\documentclass[journal, a4paper]{IEEEtran}
\usepackage{amsmath,amsfonts}
\usepackage{algorithmic}
\usepackage{algorithm}
\usepackage{array}
\usepackage[caption=false,font=normalsize,labelfont=sf,textfont=sf]{subfig}
\usepackage{textcomp}
\usepackage{stfloats}
\usepackage{url}
\usepackage{verbatim}
\usepackage{graphicx}
\usepackage{cite}
\usepackage{color}
\usepackage{soul}
\usepackage{xcolor}
\definecolor{sblue}{RGB}{0,0,0}
\newcommand\re[1]{\textcolor{sblue}{#1}}

\begin{document}
\bstctlcite{IEEEexample:BSTcontrol} 
\title{ {Inference-Driven Uplink for 6G:\\Architecture, Principles, and Challenges}
}

\author{Chunmei Xu, Yi Ma,  Rahim Tafazolli, Peiying Zhu

\thanks{C. Xu, Y. Ma and R. Tafazolli are with 5GIC \& 6GIC,
	Institute for Communication Systems (ICS), University of Surrey, Guildford,
	U.K. (emails:\{chunmei.xu; y.ma; r.tafazolli\}@surrey.ac.uk). 
}
	
\thanks{P. Zhu is with Huawei Technologies Canada, Ottawa, ON K2K 3J1, Canada
	(e-mail: peiying.zhu@huawei.com).}
}


\maketitle

\begin{abstract}
Next-generation wireless networks (6G) face a critical uplink challenge arising from stringent device-side resource constraints and the growing demand for \re{intelligent} services. 
This article introduces InferCom, an inference-driven uplink architecture designed to enable robust communication under low signal-to-noise (SNR) conditions. 
It adopts a compute-asymmetric design with a lightweight transmitter and an inference-capable receiver empowered by generative artificial intelligence models. 
Grounded in the information bottleneck principle, InferCom redefines communications through task-agnostic compression, inference-driven reconstruction, error distribution channel code, and quality of experience-aware retransmission.  A case study demonstrates that InferCom outperforms conventional 5G NR and Deep-JSCC in terms of transmitter-side computational complexity, \re{uplink coverage} and retransmission efficiency. Finally, we outline key challenges and research directions for \re{inference-driven uplink design in future intelligent 6G networks.}
\end{abstract}

\section{Introduction}
The evolution of next-generation wireless networks (6G) is being driven by intelligent applications such as immersive extended reality, telepresence, digital twins, and autonomous robots \cite{giordani2020toward}. 
These applications rely on continuous acquisition of high-dimensional sensory data, require  \re{timely, context-aware} interpretation of the physical environment, \re{placing greater emphasis on downstream task performance and user quality of experience (QoE) \cite{ImmersiveCom}}. However, conventional communication systems are designed to ensure strict bit-level fidelity rather than to preserve task-relevant semantics \cite{shi2021semantic}. This mismatch increasingly limits their effectiveness in supporting such \re{intelligent} applications. 
 
While both uplink and downlink transmissions \re{play important roles}, the sensory information \re{required by intelligent services}, including images, sensor readings and contextual observations, originates at edge devices. This makes \re{the} uplink the primary carrier of \re{intelligence-relevant} data \cite{huawei_agentverse_whitepaper}. Edge devices are often constrained in computation, bandwidth and \re{transmit power. In particular, limited transmit power restricts uplink coverage, which remains a critical challenge \cite{6GIC_whitepaper}}. By contrast, the downlink typically benefits from substantially higher transmit power, wider bandwidth, larger antenna arrays, and access to edge or cloud computing resources. Consequently, the uplink becomes the principal bottleneck in supporting intelligent 6G services.
 
\re{Although semantic communication (SemCom)  is a promising paradigm for addressing this challenge,}  \re{many learning-based approaches} rely on autoencoder-based neural networks, such as deep joint source–channel coding (Deep-JSCC) models \cite{xu2023JSCC}. \re{They typically learn an encoder to extract latent representations and a decoder to reconstruct the source while accounting for channel effects \cite{gunduz2022beyond}. While this end-to-end design is effective in principle, it imposes substantial computational and energy demands on the transmitter due to neural network execution and memory usage. Such demands often exceed the capabilities of many battery-powered or always-on sensing devices. 
In addition,  although Deep-JSCC may exhibit graceful degradation under channel noise, maintaining task-relevant reconstruction under challenging uplink conditions remains difficult, such as low signal-to-noise ratios (SNRs) or severe Doppler shifts in non-terrestrial networks \cite{xiao2024LEO}. Moreover, since end-to-end performance is often tied to the training setup, its ability to generalize across different channel conditions and source statistics remains limited.}  

\re{These limitations reveal a fundamental gap in existing solutions and call for a new architectural direction that is better suited to challenging uplink conditions, while shifting complexity away from the transmitter. Recent advances in generative artificial intelligence (GenAI), especially large-scale diffusion models \cite{ho2020denoising,  Yu_2024_CVPR, li2025diffvsr}, make such a shift increasingly attractive.}  Trained on large-scale datasets, these GenAI models exhibit powerful generative priors and can generate plausible high-fidelity reconstructions from distorted and/or degraded inputs.   Inspired by this, we introduce InferCom,  an inference-driven uplink communication architecture in which the GenAI \re{model} is deployed at the receiver. \re{ It follows a modular, compute-asymmetric design that fundamentally differs from existing autoencoder-based SemCom designs.}

\re{Building on this receiver-side inference capability, InferCom realizes the architectural shift through a lightweight transmitter, an inference-capable receiver, and a QoE-aware retransmission mechanism.} The lightweight transmitter applies simple, task-agnostic compression, considerably \re{reducing} the computational burden on edge devices.  \re{A large-scale GenAI model deployed at the resource-rich receiver serves as the inference} engine for scene understanding, semantic interpretation, and QoE-satisfying reconstruction.   By transmitting inference-critical cues, InferCom allows the receiver to recover task-relevant content even when the received observations are degraded, making it well suited for low-SNR conditions. In addition, InferCom introduces a QoE-aware retransmission mechanism, requesting additional information only when the reconstruction fails to achieve semantic adequacy. This avoids unnecessary retransmissions, which is especially beneficial in low-SNR regimes, where conventional \re{cyclic} redundancy check (CRC)-based retransmissions can become frequent and inefficient. Overall, InferCom preserves task-relevant semantics with minimal transmitter-side complexity and enables more robust and efficient uplink communication under \re{challenging channel} conditions.  

\re{The remainder of this article presents the InferCom architecture, identifies its key design principles, discusses an illustrative case study, and outlines its key challenges and research directions.}

\begin{figure*}[tp]
	\centering
	\includegraphics[width=0.8\textwidth]{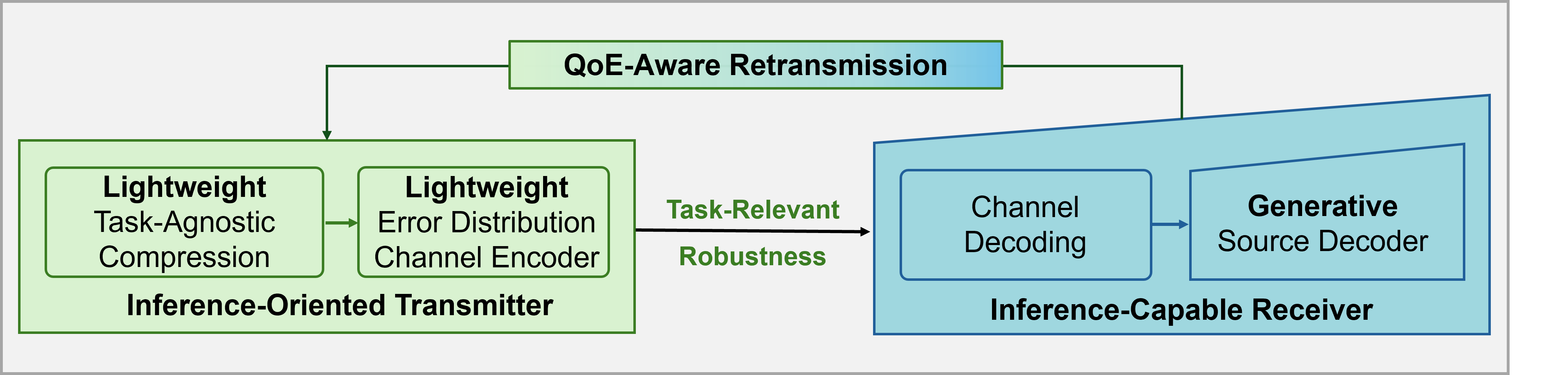}

	\caption{The proposed InferCom architecture following a modular, compute-asymmetric design.}
	\label{fig:framework}\end{figure*}
	 
\section{InferCom Architecture \label{sec:II}}
\re{InferCom aims to reliably support downstream intelligent services over challenging wireless uplinks in the 6G era, rather than learning task-specific representations for downstream tasks. This section presents the functional modules of the InferCom architecture as illustrated in Fig.~\ref{fig:framework}}. 
	
\subsection{Inference-Oriented Transmitter}
The transmitter in InferCom is inference-oriented. It produces coarse yet informative representations that retain inference-critical structural and statistical cues required by GenAI models for task-relevant reconstruction at the receiver. 
The transmitter design follows three objectives. First, it minimizes processing complexity, ensuring feasibility for compute- and energy-constrained devices. Second, it retains sufficient inference-critical information through lightweight, task-agnostic compression. \re{Third, it maintains robust transmission under power- and bandwidth-limited conditions.}
In general, any lightweight transformation that preserves inference-critical cues for task-relevant inference can serve as an inference-oriented mapping.

\subsection{Inference-Capable Receiver}
The receiver in InferCom is inference-capable and reconstructs outputs that satisfy the QoE requirement \re{for the downstream task.}  
GenAI models are deployed as generative source decoders. Their specific choice depends on both the sensed modality and the downstream task. 
\re{In a wireless system, the received representations are generally incomplete due to  both compression distortion and degradation caused by channel impairments. By leveraging generative priors on the structure and statistics of natural data}, the receiver can infer missing details and recover task-relevant content from such incomplete representations. \re{While large-scale GenAI models introduce non-negligible inference latency, the actual latency is strongly hardware-dependent and may be reduced through hardware acceleration at the network side, where compute and energy resources are more abundant.}

\subsection{QoE-Aware Retransmission Mechanism}
InferCom adopts a QoE-aware retransmission mechanism in which feedback decisions are driven by QoE. 
In conventional systems, retransmissions are triggered by CRC failure, enforcing strict bit-level accuracy. \re{However, this} may be unnecessary for the downstream task  and often  results in excessive retransmissions in low-SNR, bandwidth-limited, or \re{high-mobility} scenarios. In InferCom, \re{the receiver instead evaluates the reconstructed content using QoE metrics to determine whether retransmission is necessary for the downstream task.} If the QoE requirement is met, no retransmission is requested despite the presence of channel errors. Otherwise, the receiver requests incremental refinements that improve task-relevant quality.  
 
\subsection{\re{Architectural Comparison}}
\re{InferCom differs fundamentally from both conventional communication systems and Deep-JSCC in its transmitter design, receiver functionality, and feedback mechanism. Conventional systems rely on separate source coding, channel coding, and CRC-based feedback to ensure reliable bit delivery and high-fidelity reconstruction. Deep-JSCC replaces this modular pipeline with an end-to-end learned encoder–decoder architecture, but requires substantial neural processing at the transmitter and often suffers from SNR or channel model mismatch. In contrast, InferCom uses lightweight processing to preserve inference-critical cues, while shifting task-relevant reconstruction to a GenAI-enabled receiver with strong generative priors. As a result, InferCom changes the role of the receiver from a conventional decoder to an inference engine. This compute-asymmetric design provides a flexible framework for challenging uplink scenarios, although its applicability across modalities still depends on the underlying data structure and availability of suitable  generative priors.
}



\section{Theory and Design Principles of InferCom}

\begin{figure*}[tp]
	\vspace{1em}
	\centering
	\includegraphics[width=0.85\textwidth]{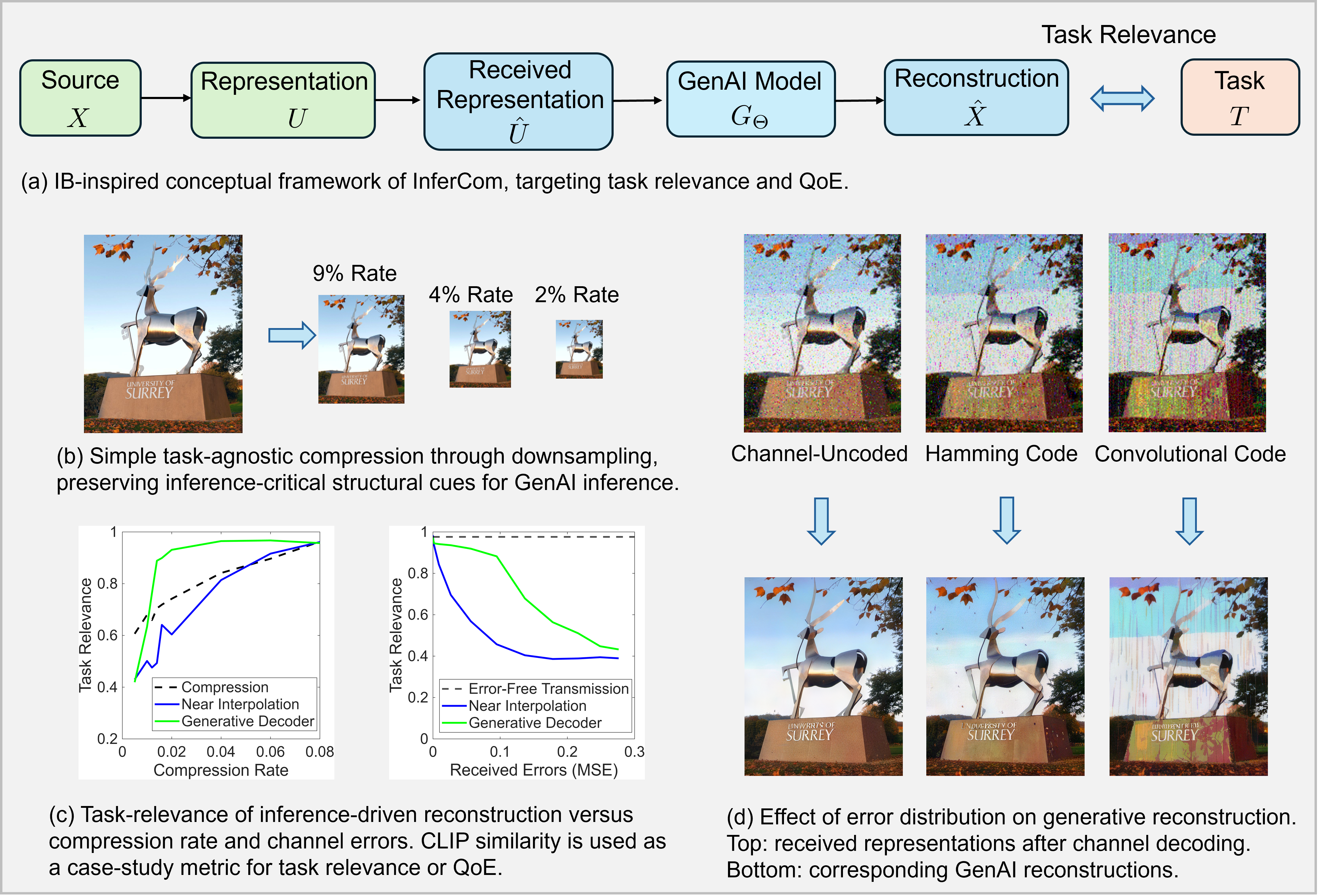}
	\caption{\re{Theory and design principles of InferCom.}}
	\label{fig:key_design}
\end{figure*}
This section explains the design principles behind the architectural modules of InferCom. These principles are grounded in the information bottleneck (IB) principle \cite{Tishby1999IB}. 
 \re{ As illustrated in Fig. \ref{fig:key_design}(a), the system aims to preserve in the transmitted representation the information that remains relevant to the downstream task, while relying on the GenAI model at the receiver to perform task-relevant reconstruction.} Here, task relevance \re{refers to the usefulness of the preserved information for the downstream task.} InferCom gives rise to four key design shifts.   
\subsection{Simple, Task-Agnostic Compression }
The first design shift concerns how the source is processed at the transmitter. Conventional compression and current deep-JSCC schemes are largely rooted in rate-distortion theory, aiming to maximize representation efficiency while ensuring \re{high fidelity}.  The resulting representations are highly compact and exhibit strong inter-symbol dependencies, making them vulnerable to channel errors. 
\re{In contrast,  InferCom adopts a task-agnostic compression principle via simple transformations, such as mean filtering and downsampling. This principle relies on inherent temporal, spatial, and/or spectral structures within the source.}
 As an example shown in Fig. \ref{fig:key_design}(b), such processing preserves inference-critical structures and cues for subsequent GenAI inference.   
 These operations \re{reduce} the mutual information between source $X$ and its representation $U$, without knowing the downstream task $T$. By exploiting the generative prior \re{$G_{\Theta}$}, task relevance can still be maintained, as compared with interpolation-based reconstruction as shown in Fig.~\ref{fig:key_design}(c). This design \re{keeps the transmitter lightweight, energy-efficient,  and task-agnostic}, which is highly practical for resource-constrained devices in 6G uplink scenarios.

\subsection{Inference-Driven Reconstruction}
The second design shift concerns how the source is reconstructed at the receiver. \re{Traditional source decoders are primarily designed to achieve high-fidelity source reconstruction, whereas Deep-JSCC decoders aim to reconstruct the source directly from noisy channel observations with minimal end-to-end distortion.}
The former relies on accurate recovery of transmitted representations, while the latter often \re{suffers from SNR mismatch or distribution shift relative to the training setup. InferCom departs from both by treating reconstruction as an inference problem. 
Under the IB viewpoint, the goal is not to reproduce the source $X$ faithfully, but to infer task-relevant contents that can meet the downstream task requirements. 
In this sense, InferCom shifts the reconstruction objective from source fidelity to task relevance, where the generative prior $G_{\Theta}$ enables the receiver to infer missing details from incomplete observations. This inference-driven reconstruction paradigm enables InferCom to support robust uplink operation with minimal transmitter-side processing under challenging channel conditions.}


\subsection{\re{Error Distribution} Channel Code}
The third design shift concerns how channel errors are handled. Conventional communications rely on strong forward error correction (FEC) codes to protect transmitted representations against fading, noise and \re{interference, typically requiring sufficiently high operating SNRs.} From the IB perspective, strict bit correctness is not always necessary for the downstream task.  InferCom instead requires received representations to retain sufficient inference-critical information for GenAI inference. 
\re{Accordingly, channel errors need not be fully removed. Their impact depends on their distribution in the received representation.}  As illustrated in Fig. \ref{fig:key_design} (d),  GenAI models are generally far more capable of handling unstructured artifacts than structured ones. This gives rise to an \re{error distribution} coding principle, which intentionally \re{shape} errors in an inference-friendly manner. 

\subsection{QoE-Aware Retransmissions}
The fourth design shift lies in the retransmission mechanism.  \re{Instead of relying on CRC outcomes as in conventional systems, InferCom requests retransmission only when the received representation is insufficient to achieve the target level of task adequacy through inference. This shift is motivated by the fact that imperfect reception does not necessarily imply inference failure.}  Under the IB lens, the retransmission is warranted  only when the additional packet is expected to provide inference-critical information that improves task relevance. 
If errors can be identified, selective retransmission of semantically important corrupted packets may be employed, where semantic importance \re{should be defined with respect to the downstream task}. 
A key design challenge is to develop  low-overhead mechanisms for assessing the semantic sufficiency without invoking full generative decoding, which may otherwise incur excessive processing delays. \re{More generally, the resulting QoE feedback can support adaptive communication loops, guiding not only retransmission decision but also the physical-layer parameter adjustments according to the required task adequacy.}



\begin{figure*}[tp]
	\vspace{1em}
	\centering
	\includegraphics[width=0.85\textwidth]{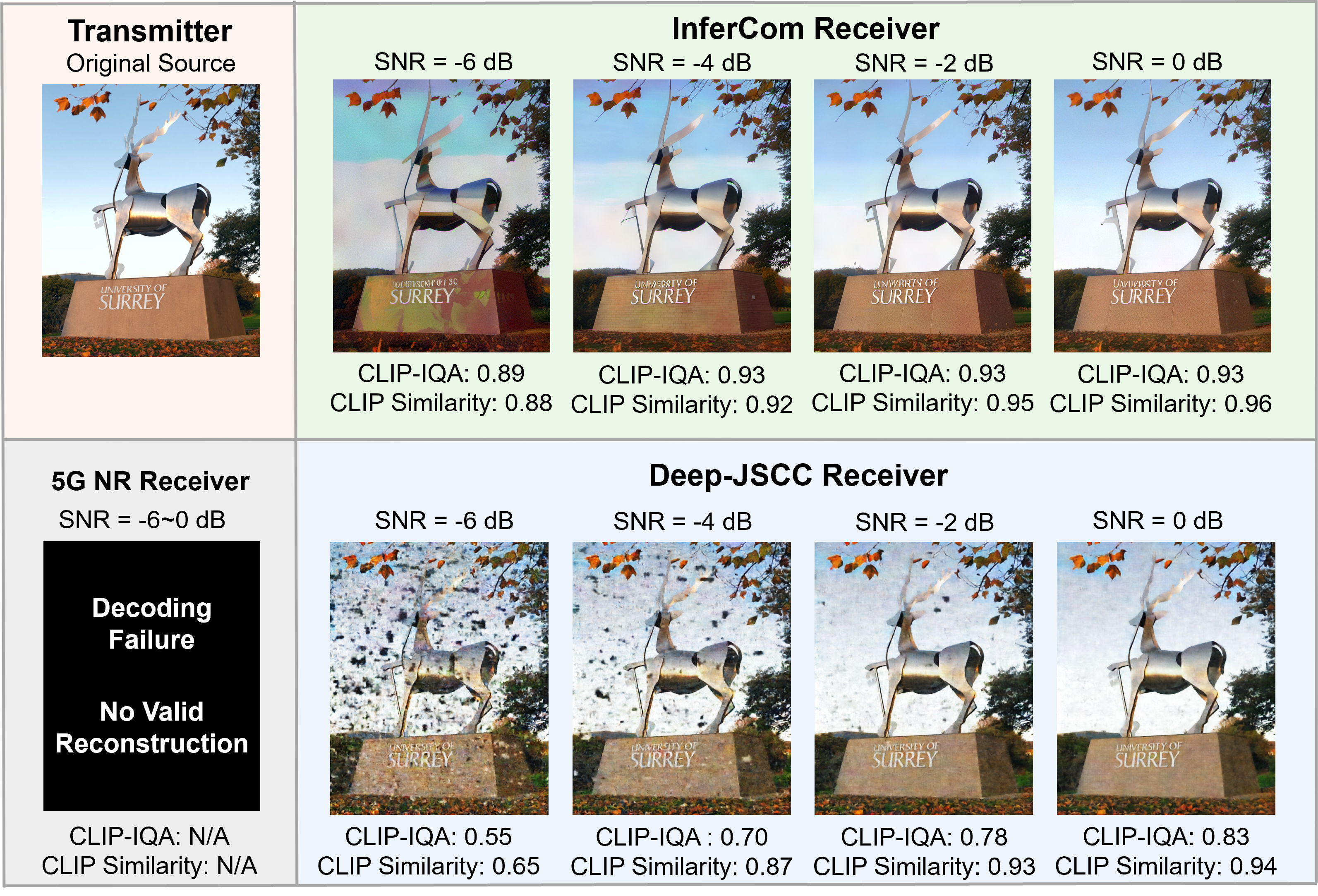}
	\caption{\re{Illustrative results at the compression rate of $0.09$ under low-SNR conditions. 5G NR fails to produce valid reconstructions  due to LDPC's poor decoding performance under low-SNR conditions. Deep-JSCC is trained at $E_b/N_o=0$ dB and evaluated across an SNR region to reflect practical channel variability}.} 
	\label{fig:results_1}
\end{figure*}

\begin{figure*}[tp]
	\vspace{1em}
	\centering
	\includegraphics[width=0.98\textwidth]{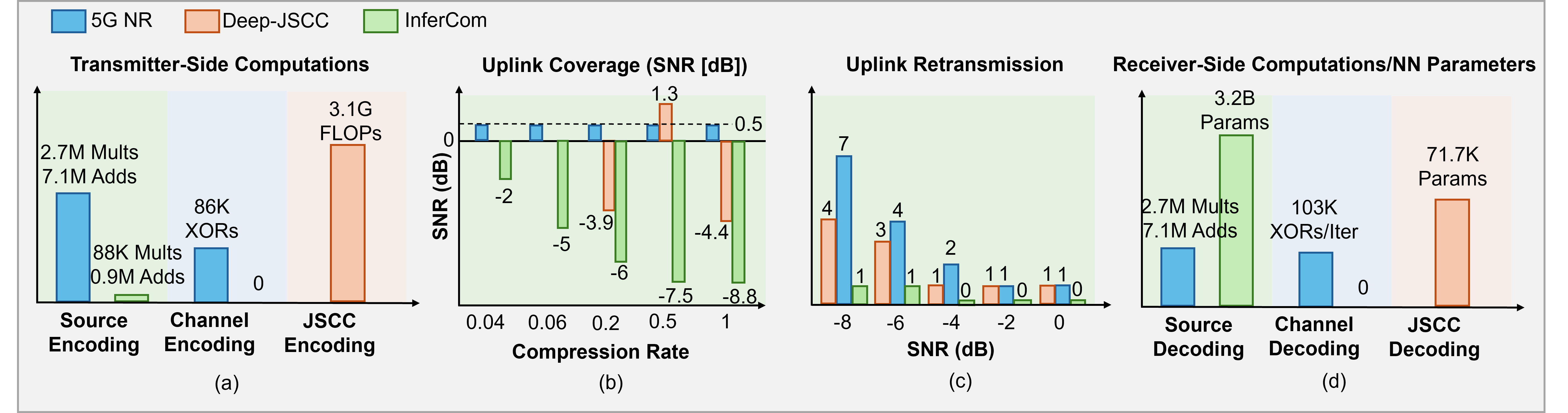}
	\caption{\re{Performance comparison: (a) Transmitter-side computational complexity. (b) SNR gains across different compression rates. (c) Retransmission rates across low-SNR settings. (d) Receiver-side computational complexity.  For 5G NR baseline, the computation for JPEG encoding/decoding is approximated by the dominated DCT/IDCT. The standardized QC-LDPC code with BG1 is adopted with the information block length of $8448$ and the coding rate of $1/3$. Deep-JSCC is trained across different compression rates at $E_b/N_o=0$ dB.
	} }
	\label{fig:results_2}
\end{figure*}

\section{Case Study and Illustrative Results \label{sec:IV}}
\re{This section presents a case study in an uplink scenario where a resource-constrained device transmits images to a BS under low-SNR conditions. Image data is adopted as a representative modality due to its intuitive visualization and the availability of well-established evaluation metrics. For simplicity, we consider a human-centric perceptual task that aims to guarantee the perceived visual quality and semantic consistency. Accordingly, QoE is evaluated using non-reference CLIP-IQA for perceptual quality and  CLIP similarity for semantic consistency, with requirement thresholds set to $0.8$ and $0.9$, respectively. These metrics are used here as proxies for the perceptual task, and should not be interpreted as universal measures for general downstream tasks.}


\subsection{Case Study Setup}

\re{The wireless channel is modeled as block Rayleigh fading, which captures block-wise uplink variations and enables evaluation across different instantaneous SNR values (e.g., $E_b/N_0$ from $0$ dB to $-6$ dB).} Within InferCom, the transmitter applies a mean filter for lightweight, task-agnostic compression, while the receiver employs \re{the publicly available open-source SUPIR model \cite{Yu_2024_CVPR} comprising $3.2$B parameters to infer missing details.  The uncoded QPSK scheme is used as an \re{error distribution} channel code. Importance-aware power allocation is used for sub-streams partitioned based on the sub-pixel level importance \cite{xu2025dataimportanceJ}.}
The QoE-aware retransmission protocol with Chase combining (CC) is adopted to enhance task relevance, \re{which can be interpreted as increasing the effective SNR.}

\re{To benchmark InferCom, 5G NR and Deep-JSCC are considered}. The 5G NR baseline uses JPEG for compression and \re{QC-LDPC} for channel coding, while Deep-JSCC learns a joint representation through an autoencoder trained under specific SNR and compression rate settings. Both baselines adopt QPSK modulation, treat compressed bits as equally important, and \re{employ} CRC-based retransmission \re{with} CC. \re{Although 5G NR systems typically employ HARQ with incremental redundancy (IR), CC is adopted here to maintain consistency with the resource-constrained transmitter assumption, offering reduced processing and memory overhead.}

\subsection{Illustrative Results}
Fig. \ref{fig:results_1} depicts the reconstructed images across a low-SNR range.  \re{The 5G NR baseline fails to produce valid reconstruction due to the strong inter-symbol dependencies introduced by JPEG compression and the limited LDPC decoding capability in low-SNR conditions.}
  At an SNR of $0$ dB, both InferCom and Deep-JSCC are able to deliver \re{satisfactory} quality. InferCom attains a perceptual score of $0.93$ and a similarity score of $0.96$, slightly higher than the Deep-JSCC receiver ($0.83$, $0.94$).   As the instantaneous SNR decreases, \re{the degradation of Deep-JSCC becomes more noticeable}.  
The largest contrast appears at $–6$ dB. Deep-JSCC degrades sharply with perception falling to $0.55$ and similarity to $0.65$, \re{while} InferCom reconstructs semantically meaningful content with a perception score of $0.89$ and a similarity score of $0.88$. 
Across the entire SNR range of interest, InferCom shows a slow  QoE degradation curve, maintaining significantly higher perception and semantic similarity than both baselines. This robustness arises from task-relevant structural cues at the transmitter and \re{the semantic inference engine} at the receiver.

\subsection{\re{Complexity, SNR Gains and Retransmissions}}
\re{In this case study, end-to-end latency and total system energy consumption are not used as primary metrics, since they depend strongly on GenAI model sizes, hardware platforms and implementation settings. By contrast, the performance metrics of computational complexity, SNR gains, and retransmission rate can directly capture the main architectural effects and design trade-offs, which are presented in Fig.~\ref{fig:results_2}.}

\re{The transmitter- and receiver-side computation comparisons are provided in Fig.~\ref{fig:results_2}(a) and (d), respectively. Deep-JSCC requires over $3.1$G FLOPs ($71.7$K parameters) for encoding and decoding. The 5G NR baseline involves approximately $2.7$M multiplications and $7.1$M additions for JPEG processing, together with $86$K XOR for QC-LDPC encoding and $103$K XOR operations per iteration for decoding. In contrast, InferCom employs a lightweight transmitter design, requiring only $88$K multiplications and $0.9$M additions for source encoding, without any channel coding operations. While the receiver performs computationally intensive generative inference (SUPIR with $3.2$B parameters), this design deliberately shifts complexity to the network side. As a result, InferCom reduces over $90\%$ of transmitter-side computation, making it well suited for compute-constrained devices in immersive sensing scenarios.}

The uplink coverage is quantified by the required SNR to meet the QoE requirement. As shown in Fig. \ref{fig:results_2} (b), 5G NR succeeds in reconstructing the image when SNR exceeds $0.5$ dB across different compression \re{rates} due to the cliff effect.  Deep-JSCC fails \re{to satisfy} the QoE requirement under low compression rates of $0.04$ and $0.06$,  and is unstable at higher compression rates. The required SNR is $1.3$ dB at a compression rate of $0.5$, but is $-3.9$ dB and $-4.4$ dB at compression rates of $0.2$ and $1$. InferCom consistently \re{requires} lower SNRs across the compression rates, with SNR gains increasing with the compression rate.  \re{This trend highlights} the receiver's inference capability to jointly compensate for both compression distortion and channel errors, and also reflects the interdependency between source and channel coding.

As shown in Fig. \ref{fig:results_2} (c), InferCom’s QoE-aware retransmission protocol further enhances uplink efficiency. Relying on strict CRC checks, 5G NR triggers the most retransmissions, followed by the Deep-JSCC and InferCom systems. Both 5G NR and Deep-JSCC request retransmission when the SNR falls below $0$ dB in contrast to $-6$ dB within InferCom. At an SNR of $–8$ dB,  InferCom reduces retransmissions by approximately $85\%$  and $75\%$ compared to 5G NR and Deep-JSCC, \re{respectively}. This \re{can translate into  substantial savings in uplink retransmission latency}.

\subsection{Key Observations and Insights}

\re{While absolute performance depends on the selected GenAI model, its primary role is to provide a strong generative prior for inference-driven reconstruction.  The following observations reflect the architectural rationale of InferCom rather than the choice of a specific GenAI model}.
First, retaining only low-frequency structural cues can be sufficient to preserve task relevance. 
Second, the generative \re{prior} of the GenAI model \re{enables QoE-satisfying reconstructions even when the received representations are incomplete or severely degraded. Third, InferCom is less tied to specific SNR conditions and channel types than Deep-JSCC, therefore potentially generalizing more effectively across diverse channel conditions.}
 Finally,  the QoE-aware retransmission \re{mechanism improves} the uplink efficiency by reducing unnecessary transmissions.

\re{The task-agnostic compression principle is most natural for modalities that exhibit exploitable temporal, spatial, or spectral structure and for which strong generative priors are available. These assumptions are well supported for natural images and are promising for audio and video, but less established for modalities such as radar point clouds or haptic signals. The extension of InferCom to multimodal settings also remains to be investigated.} 

\re{In high-mobility and wideband scenarios, doubly selective channels introduce both ISI and ICI, resulting in correlated distortion across adjacent symbols and subcarriers. Although these effects are not explicitly modeled in the present case study, they can in practice be mitigated by conventional techniques such as equalization and interleaving, and can be approximately represented through an effective post-equalization residual interference-plus-noise model \cite{Wang2012}. As such, the qualitative insights underlying InferCom are expected to remain applicable.}

\section{Key Challenges and Future Directions \label{sec:V}}

\subsection{\re{Semantic QoE Metrics}}
\re{Modeling semantic QoE is challenging, as it must be task-specific to reflect the requirements of downstream intelligent services. Furthermore, the unavailability of original sources at the receiver makes non-reference semantic QoE metrics both essential and non-trivial.} Meanwhile, traditional QoS indicators, such as BLER and SINR, are directly used to determine modulation and coding schemes, \re{beamforming}, retransmissions, and scheduling decisions, but they do not capture semantic adequacy. 
\re{Therefore, principled mappings from semantic QoE to PHY/MAC-layer QoS are needed. Such mappings should incorporate semantic importance and support task-aware lower-layer designs. Specifically, they allow the network to translate  QoE requirements into target QoS levels, based on which PHY/MAC-layer parameters can be adaptively configured.} 


\subsection{A Novel Protocol Stack and Cross-Layer Design}
The communication protocol architecture introduces another fundamental challenge. Existing wireless systems are built on a layered protocol stack. The application layer handles compression/reconstruction or semantic encoding/decoding.  The transport/MAC layers manage flow control and retransmissions.  The PHY layer decides modulation, coding, and waveform adaptation. This separation prevents the lower layers from exploiting semantic redundancy or semantic importance for more effective modulation, coding, scheduling decisions, and radio resource optimization. InferCom challenges this architecture by requiring a new cross-layer protocol stack. In this design, semantic QoE guides link adaptation, \re{HARQ} decisions, and resource allocation at lower layers. Semantic importance and the receiver's inference capability can then be leveraged to enhance the system's robustness. Developing such a protocol stack is non-trivial and calls for rethinking long-standing design boundaries between layers, redefining signalling interfaces, and redesigning control loops to support inference-driven communication.

\subsection{GenAI Inference Energy \re{and Latency} Modeling}
While the transmitter is intentionally lightweight, the receiver incurs substantial energy consumption \re{and latency} due to large-scale generative inference, making accurate modelling essential for InferCom. Energy modeling requires a comprehensive framework \cite{faiz2024llmcarbon} that \re{captures} parameter-related memory footprint, FLOP-based computational complexity, hardware-aware execution efficiency (e.g., memory hierarchy and throughput bottlenecks), and system-level overheads (e.g., cooling and platform-specific execution).  \re{Meanwhile, inference latency depends not only on raw computational workload, but also on memory access and system-level scheduling or queueing effects, which may become dominant bottlenecks in large-scale generative models. These two aspects are inherently coupled, leading to an energy–latency tradeoff that is critical for the design of GenAI models in 6G systems. This tradeoff becomes more pronounced when supporting multiple concurrent inference tasks, posing additional challenges in compute resource allocation and inference scheduling.}  
 
\subsection{\re{Multimodal Multiple Access}}

\re{
InferCom shows robustness to transmission errors in the single-user setting through inference-driven reconstruction. Extending this capability to multi-user scenarios, however, introduces new challenges, as interference may degrade inference performance.
A key open problem is to characterize the relationship between inference performance and interference conditions and to identify the operating regimes in which degradation becomes significant. Such understanding will provide useful guidance for interference management,  particularly in dense deployments. Furthermore, when multiple users transmit sensory data for a shared downstream service, the problem becomes multimodal fusion under wireless constraints. In this context, InferCom may support multimodal fusion by enabling joint processing of heterogeneous observations at the receiver. Achieving this, however, requires future multiple access designs to further account not only for interference, but also for cross-modal correlation across distributed sources and possible misalignment among the received modalities.}

\subsection{Security and Semantic Inference Attacks}
The ability of InferCom to operate under extremely low-SNR conditions with high semantic fidelity faces a new class of security and semantic inference attacks. An adversary equipped with a powerful GenAI model may infer transmitted content even from severely degraded signals. This challenges traditional PHY-layer security schemes that are grounded \re{in} secrecy capacity. 

\section{Conclusion  \label{sec:VI}}
\re{This article introduced InferCom, an inference-driven uplink framework for 6G, motivated by the resource constraints of edge devices and the growing demand for intelligent services. With its compute-asymmetric design, InferCom departs from both conventional bit-centric systems and learned end-to-end SemCom paradigms through four key design shifts. 
The case study suggests that InferCom can substantially reduce transmitter-side complexity while improving robustness and retransmission efficiency under low-SNR conditions, by shifting the main inference burden to the receiver side. Realizing this vision will require further advances in semantic QoE modeling, cross-layer protocol design, inference energy and latency modeling, multimodal multiple access, and security.}

\bibliographystyle{IEEEtran}
\bibliography{reference}
\vspace{-12pt}
\begin{IEEEbiographynophoto}{Chunmei Xu} (chunmei.xu@surrey.ac.uk) is currently a Research Fellow at the Institute for Communication Systems, University of Surrey, Guildford, U.K. 
Her research interests are AI-native semantic communications and sensing. 
\end{IEEEbiographynophoto}

\begin{IEEEbiographynophoto}{Yi Ma} (y.ma@surrey.ac.uk)  is currently a Chair Professor at the Institute for Communication Systems, University of Surrey, Guildford, U.K. He is the Head of Artificial Intelligence for Wireless Communication Group within the ICS to conduct the fundamental research of wireless communication systems covering signal processing, applied information theory and artificial intelligence. 
\end{IEEEbiographynophoto}

\begin{IEEEbiographynophoto}{Rahim Tafazolli} (r.tafazolli@surrey.ac.uk)  is Regius Professor of Electronic Engineering, Professor of Mobile and Satellite Communications, Founder and Director of 5GIC, 6GIC and ICS (Institute for Communication Systems) at the University of Surrey.  He holds Fellowships of the Royal Academy of Engineering, the Institute of Engineering and Technology (IET), as well as that of Wireless World Research Forum. He was also awarded the 28th KIA Laureate Award-2015 for his contribution to communications technology. 
\end{IEEEbiographynophoto}

\begin{IEEEbiographynophoto}{Peiying Zhu} (peiying.zhu@huawei.com)  is Fellow of the Canadian Academy of Engineering. She is currently leading 6G wireless research at Huawei Technologies, Canada. She is actively involved in 3GPP and IEEE 802 standards development. The focus of her research is advanced radio access technologies with more than 150 granted patents.
\end{IEEEbiographynophoto}


\vfill

\end{document}